%% file: paper_reportingManifesto.tex
\documentclass[twoside,twocolumn]{article}
\usepackage{blindtext} % Package to generate dummy text throughout this template 

\usepackage{cite}
\usepackage[utf8]{inputenc} 
\usepackage[sc]{mathpazo} % Use the Palatino font
\usepackage[T1]{fontenc} % Use 8-bit encoding that has 256 glyphs
\usepackage{microtype} % Slightly tweak font spacing for aesthetics

\usepackage[hmarginratio=1:1,top=32mm,columnsep=20pt]{geometry} % Document margins
\usepackage[hang, small,labelfont=bf,up,textfont=it,up]{caption} % Custom captions under/above floats in tables or figures
\usepackage{booktabs} % Horizontal rules in tables

\usepackage{lettrine} % The lettrine is the first enlarged letter at the beginning of the text

\usepackage{enumitem} % Customized lists
\setlist[itemize]{noitemsep} % Make itemize lists more compact

\usepackage{abstract} % Allows abstract customization
\usepackage{titlesec} % Allows customization of titles
\titleformat{\section}[block]{\large\scshape\centering}{\thesection.}{1em}{} % Change the look of the section titles
\titleformat{\subsection}[block]{\large}{\thesubsection.}{1em}{} % Change the look of the section titles

\usepackage{graphicx}
\usepackage{subfig}
\usepackage{fancyhdr} % Headers and footers
\usepackage{titling} % Customizing the title section

\usepackage{hyperref} % For hyperlinks in the PDF

\usepackage{amsmath} % For math typesetting, in particular the \begin{aligned} environment to have multiline equations

\usepackage{pifont}% http://ctan.org/pkg/pifont

\usepackage{xcolor}

\usepackage{mathtools}

\newif\ifpaper
\papertrue

\input{newcommands}

\pretitle{\begin{center}\Huge\bfseries} % Article title formatting
\posttitle{\end{center}} % Article title closing formatting
\title{Reporting Results in High Energy Physics Publications: a Manifesto} % Article title
\author{%
  \textsc{Pietro Vischia}\\[1ex] % Your name
\normalsize Institut de recherche en Math\'ematique et Physique,\\ % Your institution
\normalsize Universit\'e catholique de Louvain\\
\normalsize \href{mailto:pietro.vischia@cern.ch}{pietro.vischia@cern.ch} % Your email address
}
\date{\today} % Leave empty to omit a date
\renewcommand{\maketitlehookd}{%
\begin{abstract}
  \noindent The complexity of collider data analyses has dramatically increased from early colliders to the CERN LHC.

  Reconstruction of the collision products in the particle detectors has reached a point that requires dedicated publications documenting the techniques,
  and periodic retuning of the algorithms themselves.
  Analysis methods evolved to account for the increased complexity of the combination of particles required in each collision event (final states)
  and for the need of squeezing every last bit of sensitivity from the data;
  physicists often seek to fully reconstruct the final state,
  a process that is mostly relatively easy at lepton colliders but sometimes exceedingly difficult at hadron colliders to the point of requiring
  sometimes using advanced statistical techniques such as machine learning.

  The need for keeping the publications documenting results to a reasonable size implies
  a greater level of compression or even omission of information with respect to publications from twenty years ago.
  The need for compression should however not prevent sharing a reasonable amount of information that is essential to understanding a given analysis.
  Infrastructures like \textsc{Rivet} or \textsc{HepData} have been developed to host additional material,
  but physicists in the experimental Collaborations often still send an insufficient amount of material to these databases.
  
  In this manuscript I advocate for an increase in the information shared by the Collaborations,
  and try to define a minimum standard for acceptable level of information when reporting the results of statistical procedures in High Energy Physics publications.
\end{abstract}
}

\begin{document}

% Print the title
\maketitle

%----------------------------------------------------------------------------------------
%	ARTICLE CONTENTS
%----------------------------------------------------------------------------------------
\input{content_paper}

\section*{Acknowledgements}
I wish to thank the ATLAS and CMS Collaborations for being an endless source of inspiration,
the CMS Collaboration for having me among its members since exactly ten years, four of which within the Statistics Committee, and the SM@LHC 2019 participants:
the organizers for hosting the talk that inspired me,
and all the participants for the stimulating discussions that helped me refine the ideas behind this manuscript. 

A special thanks goes to: Pedro Silva and Tommaso Dorigo for their useful comments on an earlier draft of this manuscript;
Fabio Cossutti for a nice conversation on reporting unfolding results, while walking along the banks of the Z\"urichsee;
and to Javier Cuevas for the endless discussions and encouragement that helped me shape my own critical views on a broad selection of topics.

All the feedback I received contributed to largely improving this manuscript.
Benjamin Fuks, Andrea Giammanco, Alexander Held, Patrick Koppenburg, Alessia Saggio, Nick Wardle, and Graeme Watt sent very useful feedback on earlier versions: I am very grateful to them.

This work was supported by the Universit\'e catholique de Louvain, Belgium, under the FSR grant \texttt{IRMPFSR19MOVELem}.

%----------------------------------------------------------------------------------------
%	REFERENCE LIST
%----------------------------------------------------------------------------------------

%\printbibliography
%[heading="References"
\bibliographystyle{splncs03_unsrt}
\bibliography{bibliography}
%\bibliographystyle{ieeetr}
% wget https://jhep.sissa.it/jhep/help/JHEP/TeXclass/DOCS/JHEP.bst

%----------------------------------------------------------------------------------------

\end{document}

%% file: newcommands.tex
\newcommand{\pdf}{p.d.f.}

\newcommand{\fbinv}{\ensuremath{\mathrm{fb^{-1}}}\xspace}

%% file: content_paper.tex
\section{Introduction}

\begin{figure}[!t]
  \centering
  \includegraphics[width=\linewidth]{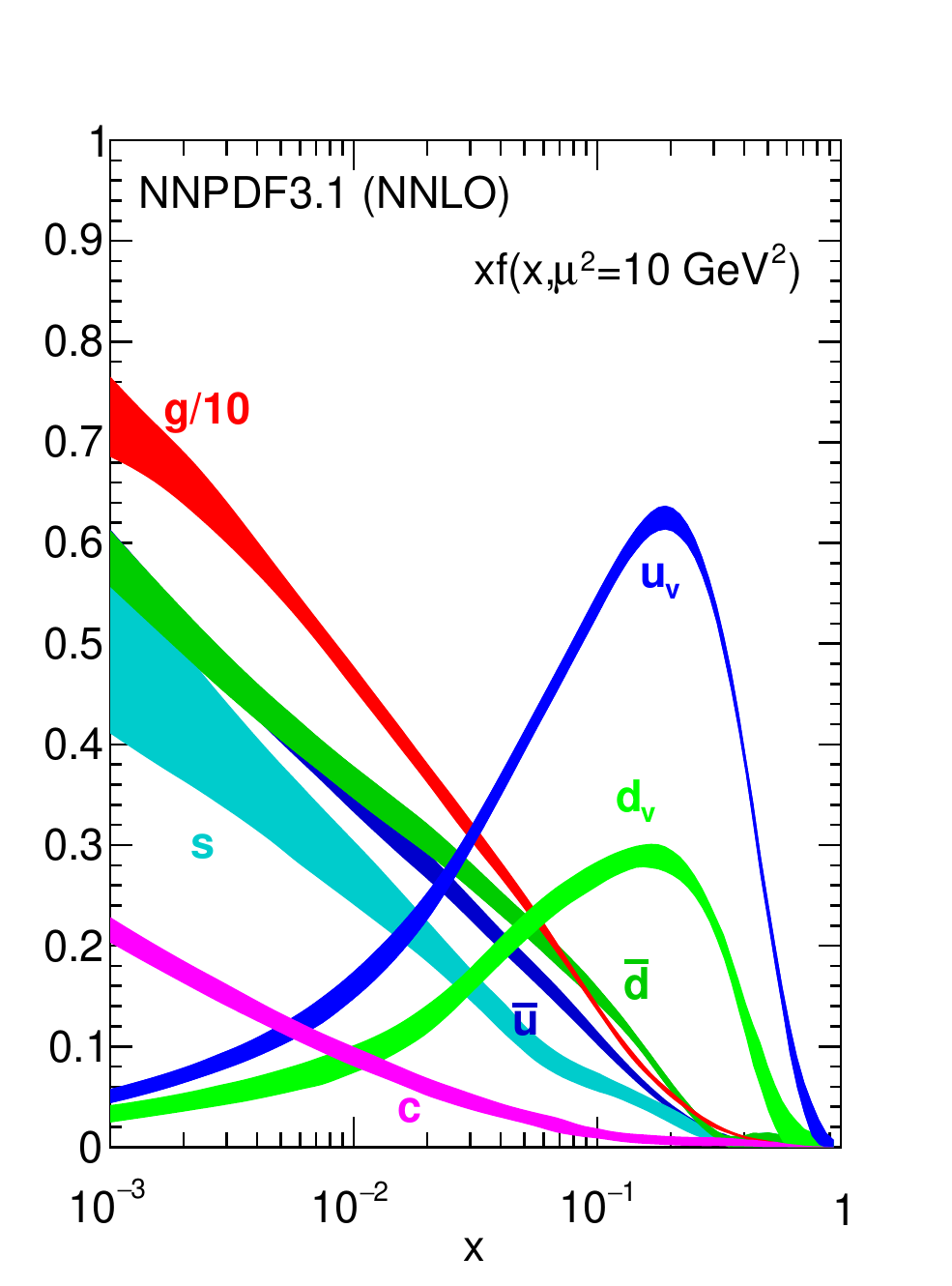}
  \caption{The NNPDF3.1 next-to-next-to-leading-order (NNLO) PDFs, computed at a scale $\mu^2=10~{\rm GeV}^2$ by the NNPDF Collaboration~\cite{Ball:2017nwa}}
  \label{fig:nnpdf31}
\end{figure}

Analyses of collider data have become increasingly more complex all the way from early colliders to the CERN LHC.
Earlier lepton colliders like LEP yielded very clean final states, which favoured simple reconstruction and analysis methods;
such results were easily reported in detail even in brief publications.
The advent of hadron colliders such as the Fermilab Tevatron introduced a new level of complexity.
In lepton collisions, the momentum in the longitudinal direction with respect to the particles beam is precisely known;
In hadron collisions, protons or antiprotons are accelerated in bunches where only the average momentum $\langle p\rangle\pm \Delta\langle p\rangle$ is controlled;
more importantly, when two hadrons collide, the actual collision is between quark constituents carrying only a fraction of the hadron momentum~\cite{Soper:1996sn}.
We know only the probability distribution of the momentum fraction---the \textit{parton distribution functions} (PDFs)---which we determine from global fits involving all known theoretical and experimental constraints; an example is the NNPDF-3.1 set (Figure~\ref{fig:nnpdf31}).

Even if we can partially account for the PDFs when generating simulated events, we have no experimental access to a reliable estimate of
the longitudinal momentum $p_z$ of the individual colliding partons. The unaccessibility of $p_z$ forces us to reconstruct
quantities mainly in the transverse plane, where momentum conservation assuming a perfectly longitudinal beam tells us that the sum of all the momenta
of the collision products must be zero; when the sum of the momenta of the detected particles is not zero, we infer that the negative
sum of the tranverse momenta describes the transverse component of the energy of whichever invisible decay product---neutrinos, in the Standard Model of Particle Physics~\cite{Glashow:1961tr,Weinberg:1967tq,Salam:1968rm,GellMann:1964nj,Zweig:570209,Fritzsch:1973pi}---are produced.
As a consequence, in hadron colliders we can fully reconstruct final states easily only when we expect that the missing transverse energy
can be assigned to a single neutrino, i.e. when the scattering process in that final state involves only one neutrino;
even then, instrumental effects in the determination of the transverse energy may deteriorate the full event reconstruction capability
in a non-trivial way. When we expect more that one neutrino in the final state, we need to use more advanced
techniques---for example, accounting for inaccuracies in the measured quantities that enter the reconstruction equations~\cite{Betchart:2013nba} due to instrumental effects.

Because we accelerate particles in bunches, final states at hadron colliders are also characterized by additional ``pileup'' interactions on top of the hard scattering,
resulting in dramatically higher hadronic activity in the event.
Theoretical advancements also play a big role in both limiting the sensitivity of the results and increasing the complexity of the analyses;
in order to tackle these issues it is crucial to have a full understanding of the uncertainties---particularly the theory ones---associated to the results.

Following the evolution of colliders, analyses of their data had to adapt.
The instantantaneous luminosity, that is the average number of collisions per second, has grown by several orders of magnitude in the past few decades. This has in turn increased the average number of pileup interactions in each collision, resulting in larger tracker and calorimeter occupancies: the reconstruction of particles in the detectors became therefore more complex.
The analysis methods also evolved to account for the increased complexity of the final states,
resulting for example in non-trivial assignment problems.
When studying the associated production of a Higgs boson in association with a pair of top quarks we may for example select final states with one or two leptons, and categorize the events according to the number of jets coming from a b quark~\cite{Aaboud:2017rss}. To reconstruct the Higgs boson it is necessary to disentangle the combinatorics stemming from the large number of jets in the final state that likely come from a b quark.
A far simpler case consists in studying the associated production of a $Z$ and a $W$ boson we may for example select leptonic decays $Z\to\ell\bar{\ell}$ and $W\to\ell'\nu_{\ell'}$
by requesting a final state with three leptons and missing energy; we then 
have to decide which leptons come from the leptonic decay of the $Z$ boson and which one comes from the leptonic decay of the $W$ boson~\cite{Sirunyan:2019bez}, but a constraint on the $Z$ boson mass is usually sufficient to disentangle the leptons from the $Z$ and the $W$ boson. 
Squeezing every last bit of sensitivity from the data often involves either the full reconstruction of the final state---a problem that was
mostly relatively easy at lepton colliders and is now sometimes exceedingly difficult at hadron colliders---or the use of advanced statistical techniques such as machine learning.

Twenty years ago, many publications in High Energy Physics (HEP) were only a few pages long.
Nowadays, the reconstruction and analysis techniques are often more complicated, and to keep the publications to a reasonable page count
some information is extremely compressed or even omitted.
This compression should not prevent from sharing a reasonable amount of information that is essential to understand a given analysis.

A similar lack of information can be found in literature documenting conventions relative to individual Collaborations or joint agreements.
A BaBar document on recommended statistical procedures~\cite{babar} advocates for example quoting as the norm only total uncertainties
rather than uncertainties split by source, except in some particular cases; this is an outdated position which I discuss in Section~\ref{sec:uncertainties}.

The ATLAS and CMS Collaboration documented procedures for obtaining results related to the Higgs boson search~\cite{CMS-NOTE-2011-005};
such procedures have since been used for most Higgs physics results and in other areas as well.
The joint-document section dedicated to reporting results recommends the publication of only a very minimal set of information
and leaves to the individual Collaborations any decision on eventually extending this set. 

This manuscript is inspired by a detailed comparison of the early-2019 ATLAS and CMS multiboson (W, Z) measurements
  in proton-proton collisions at the LHC with centre-of-mass energy of 13 TeV;
the observations I make are however of general nature, and could likely have been inspired by any comparison of recent HEP results.

An overview of the measurements themselves is beyond the scope of this work and can be found in Ref.~\cite{vischia_smatlhc_talk};
here I instead make an attempt at abstracting some general recommendations on executing and reporting (mostly) statistical
procedures, that I believe can help in comparing results within and across experiments and with theoretical predictions.
This manuscript is based on comparing a set of results in a given point in time; I therefore reference some of the experimental results
in the publication status they were in during the end of April 2019.
You are encouraged to check the ATLAS and CMS Public Results pages~\cite{atlaspublic,cmspublic} for the latest and greatest results.
For theoretical publications---and for more recent experimental publications---I cite, where appropriate, the latest published versions.

In this manuscript, \textit{they} refers to the authors of a referenced publication,
to indicate specific choices they made;
\textit{we} refers to us particle physicists as a community,
to indicate sets of common---official or unofficial---practices;
\textit{I} refers to me, to indicate my opinion on the many topics described below.

I will briefly describe the techniques connected
to reporting different types of results;
I will advocate for an increase in the information shared by the Collaborations;
and I will try to define a minimum standard for acceptable level of information when reporting
statistical procedures in HEP publications.

\section{Data sets and simulation: compare apples with apples}

The physics program of the LHC experiments is large,
encompassing both standard model (SM) and beyond standard model (BSM) physics;
analyses might be broadly divided into SM precision measurements
(which nowadays have reached a precision that enables probing even BSM scenarios,
mostly by measuring the couplings between SM particles)
and traditional searches for new physics (e.g. searches for supersymmetric partners~\cite{Martin:1997ns} of ordinary particles).
While ATLAS and CMS naturally try to analyze the same data sets roughly at the same time,
matching results are often made public with significant discrepancies.
It is not uncommon that the latest results on a given topic are derived with very different integrated luminosities.
An extreme example is WW production, where the ATLAS Collaboration~\cite{Aaboud:2017qkn} focused on publishing quickly
an inclusive measurement performed with about 3~\fbinv;
the CMS Collaboration~\cite{CMS-PAS-SMP-18-015} has instead focused in making public the evidence for WW production
in double-parton-scattering with 77~\fbinv.

Unpredictable delays in the publication of an analysis might influence this picture, and each Collaboration might have different priorities.
While I would advise to better synchronize in time, whenever possible, the analysis of similar data sets,
it is also true that some degree of temporal displacement might inform the development of any later batch of analyses,
until hopefully all Collaborations can put the final word with a matching data set at around the same time.
I think that the displacement should not be too large---cases like the one outlined above seem rather excessive.
It makes generic sense to set tight deadlines to obtain a first result before another Collaboration
(e.g. to publish an observation with the first collected data in a collider run),
but for all other analyses (including updates of previous analyses that differ essentially just by the amount of data
and minor changes to analysis methods) considerations on temporal placement should not contemplate fear-induced deadlines
(\textit{``otherwise they will publish it first''}). While \textit{``let's be the first in observing this process''} makes sense,
\textit{``let's be the first to have analyzed the full run data set''} certainly does not.
Other more delicate reasons may exist (freeing up the person power for other studies or future runs,
securing funding near the expiration of a project, etc.), but the extent to which they influence analysis deadlines
should not be overinflated. 

Simulated samples are sometimes different;
for the latest ZZ cross section results the ATLAS Collaboration \cite{Aaboud:2017rwm} uses \textsc{SHERPA}~\cite{Gleisberg:2008ta,Schonherr:2008av} for the nominal signal templates
and \textsc{POWHEG}~\cite{Nason:2004rx,Frixione:2007vw,Alioli:2010xd,Melia:2011tj,Nason:2013ydw} for alternative templates used for systematic uncertainties;
the corresponding analysis by the CMS Collaboration~\cite{CMS-PAS-SMP-19-001} uses \textsc{POWHEG} for the nominal signal templates.
Signal modelling is nowadays crucial for many precision measurements;
different conventions might result in serious issues in comparing results,
and might become show-stoppers for ATLAS+CMS grand combinations.
The ATLAS and CMS Collaborations have studied this effect in the combination of the Higgs boson production and decay rates~\cite{Khachatryan:2016vau},
where the two Collaborations took great care in using the same generators for the most sensitive channels,
and verified that using different generators had a negligible impact compared to that of the respective dominant uncertainties for the less sensitive channels.

One reason that partly explains this situation is that each Collaboration's framework
for event simulation implies different challenges in integrating
a given generator---at a given point in time for some physics process a Collaboration might be limited to only a subset
of the available generators---but when the same generator is available
to both Collaborations an effort should be done to uniformize the choice.
In case there are specific reasons for choosing one event generator rather than the other as nominal prediction,
one would expect that an agreement can easily been reached;
if there is no particular reason for preferring one event generator over the other, there should be no issue in agreeing either.

An argument against using the same generators might be to have the two Collaborations provide independent cross-checks of results,
but I argue that this would result in convoluting together any discrepancy caused by physics with any discrepancy caused by Monte Carlo modelling;
using the same generators decouples such effects,
and leaves always open the option of doing all the cross-checks within each Collaboration with all the generators available---either by trying
multiple generators for nominal predictions or by a detailed study of systematic uncertainties.
Non-trivial differences in the Monte Carlo production of the two experiments might also be explored by starting from the same set of hard scattering events
rather than from the same generator cards. This would also in most cases reduce significantly the computing cost of such calculations,
in particular the computation of the matrix element describing the hard scattering interaction as done e.g. by the \textsc{Madgraph} generator~\cite{Alwall:2014hca}.

\section{Display what you use, and choose the same visualizations}

Graphical displays of distributions---{\em plots}, in HEP jargon---represent mostly a visual aid
to show the agreement between data and predictions and increase the confidence
in the modelling of the relevant physics processes.
When the statistical inference leading to the result uses unbinned information only,
the choice of binning done for visualization purposes is purely illustrative
and the figure captions should make this explicit.
Oftentimes, however, the statistical procedures applied
to extract estimates for a parameter of interest take binned distributions as an input;
in this case, displaying a binned distribution has the important function of highlighting
the interpretation of the whole procedure as a counting experiment in orthogonal selections (\textit{regions}) corresponding
to the histogram bins.
It is crucial to use in the published histograms the same binning that characterizes the statistical model.
Sometimes you would be tempted to use a finer binning when displaying distributions---to show the extent
to which the distributions are sufficiently modelled---and a coarser binning when building the binned statistical model (typically to avoid unrealistic sensitivity driven by statistical fluctuations).
This might however mislead the reader into thinking that the final result
is affected by any empty bin or problematic tail fluctuation visible in the finely binned plot used for display.
Without space constraints it would be possible to publish both binnings,
but this is practically never feasible; it is therefore best to display in the plots exactly
what goes into the statistical model.
Sometimes the same concern about possible misleading visual fluctuations
might lead to change the binning for the visualization precisely to prevent
 the reader from interpreting as a possible excess some fluctuation visible in the distribution binned as in the statistical model;
I argue that in such cases it would be better to publish
the original binning and add a sentence explaining that any fluctuation is accounted for
in the statistical technique used to obtain the result.

Uncertainty bars and bands should likewise be reported clearly.
Statistics teaches us that the bin counts of a histogram follow a multinomial distribution constrained by the total number of events $N$,
which in turn can be considered as a fixed number (an observed count)
or as a Poisson random variable for a rare event produced at a constant rate
(highlighting the existence of an expected value for $N$).
James~\cite{James2ed} clarifies that the two assumptions result in different covariance matrices with different rank:
in the first case the bin counts are mutually anticorrelated, in the second one the bin counts are independent from each other but correlated with the total number of events.
Lyons~\cite{LyonsBook} points out that the product of a multinomial probability distribution for the bin counts, multiplied by
the Poisson probability for the total number of events $N$, is equivalent to the product of independent Poisson probabilities for the individual bins.
The number $n$ of events in each bin is taken as the Poisson mean; the uncertainty can therefore be estimated as $\sqrt{n}$,
the variance of a Poisson distribution of mean $n$.
When $n$ is small (approximately less than 10), the Poisson distribution is markedly asymmetric (skewed) and the expression $\sqrt{n}$
leads to quoting an uncertainty which is too optimistic.
To fix this problem, Garwood~\cite{10.1093/biomet/28.3-4.437} computed an interval
from first principles by using the Neyman construction for central intervals,
obtaining correct-coverage intervals even in the case of low counts or empty bin contents.
It is important to replace $\sqrt{n}$ with the Garwood intervals, particularly for low counts,
and to draw the uncertainty bars also for the empty bins:
the resulting error bars prevent us to be fooled by ``excesses'' of a handful events.

\begin{figure}[!t]
  \centering
  \includegraphics[width=\linewidth]{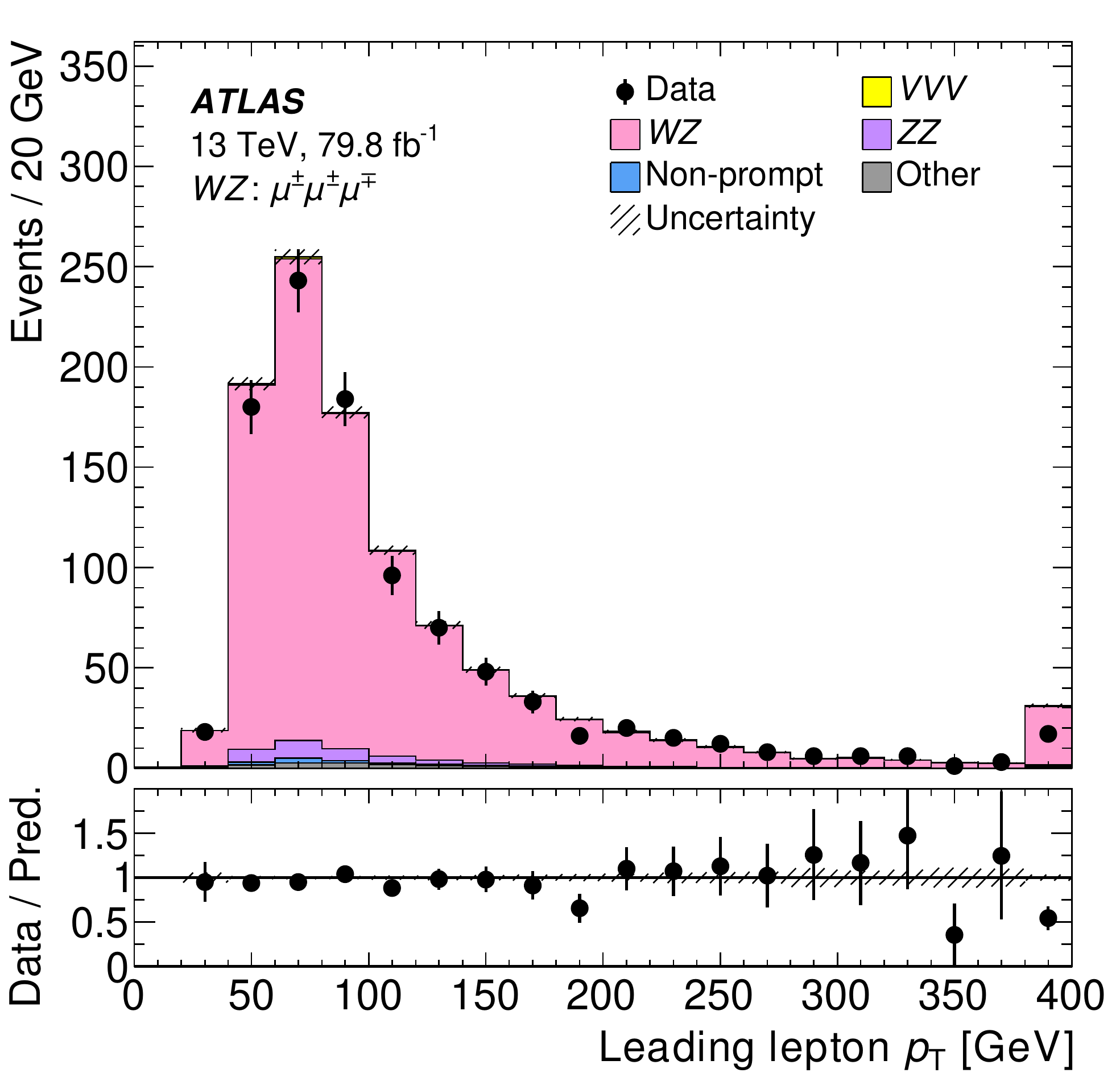}\\
  \includegraphics[width=\linewidth]{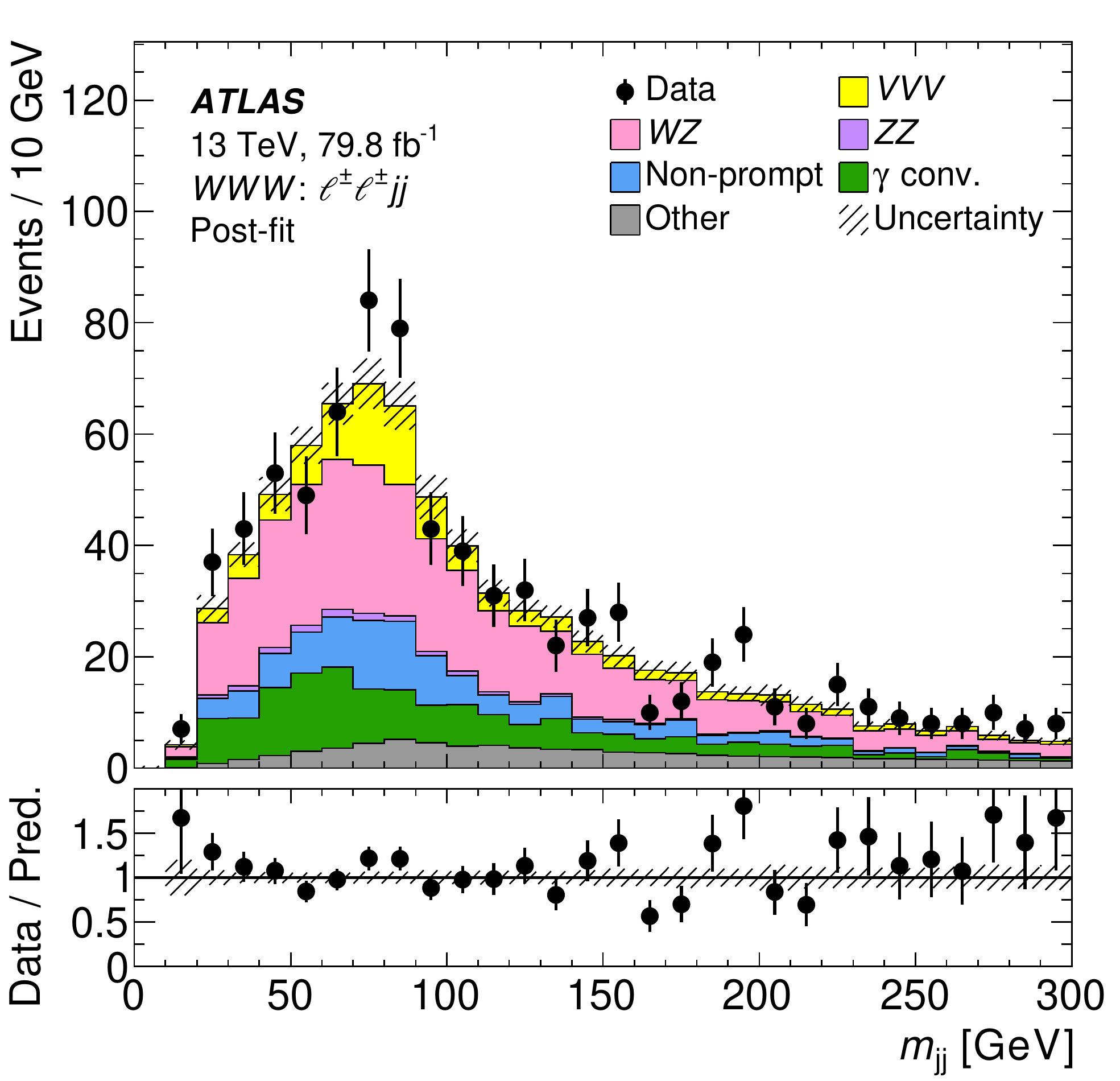}
  \caption{The leading lepton transverse momentum distribution in a validation region (top) and the dijet mass distribution in a signal region (bottom), in a publication reporting evidence for triboson production in ATLAS~\cite{Aad:2019udh} (v1). In the top figure the hatched \textit{Uncertainty} is statistical only, whereas in the bottom figure it is the squared sum of the statistical and systematic uncertainties. Later versions of the publication (v3) now explicitly mention \textit{Stat. uncertainty} in the legend of the top plot.}
  \label{fig:triboson}
\end{figure}

We must also be explicit and consistent on the content of the uncertainty bars in the predictions;
the ATLAS Collaboration reports in all figures of Ref.~\cite{Aad:2019udh} (v1) a hatched ``Uncertainty'',
and it is only by reading the caption that the reader figures out that for approximately half of the figures
the uncertainty is statistical only, while for the other half it is total (statistical plus systematic).
Figure~\ref{fig:triboson} shows an example where the hatched band represents the statistical uncertainty only (top)
and one where the hatched band represents the squared sum of the statistical and systematic uncertainties (bottom).
Later versions of that publication (v3) now explicitly mention \textit{Stat. uncertainty} in the legend, where appropriate.

While some uniformity is desirable, mixing plots with different definitions of uncertainty bands is still a reasonable option.
What is crucial---particularly for presenting a result at conferences---is that
different definitions of bands have a different pictorial representation in terms of shaded area style
and that both the legend and the caption are explicit (i.e. ``Stat. unc.'' and ``Total unc.'' instead of just ``Uncertainty''),
as is done e.g. in a recent work by the CMS Collaboration~\cite{Sirunyan2019}.

The choice of the physics observable to be reported is sometimes incoherent across Collaborations
and even within the same Collaboration.
The ATLAS and CMS Collaborations parameterize the measurements of the production cross section of associated $WZ$ production
also as a function of the electric charge of the $W$ boson.
The ATLAS Collaboration~\cite{Aaboud:2019gxl} reports the ratio $\sigma(W^{+}Z)/\sigma(W^{-}Z)$,
whereas the CMS Collaboration~\cite{Sirunyan2019} reports the ratio $A_{WZ,obs}^{\pm}/A_{WZ,NLO}^{\pm}$ of the observed ratio
$A_{WZ,obs}^{\pm} = \sigma(W^{+}Z)/\sigma(W^{-}Z)$ to the corresponding NLO expected ratio
$A_{WZ,NLO}^{\pm}$.
This is by itself an issue: the reader cannot compare the raw numbers and must instead convert results from an observable to another,
raising issues like how to correctly propagate the systematic uncertainties.
Some results, e.g. on anomalous couplings in Effective Field Theory (EFT), can be computed and reported in terms of different operator sets (bases):
while it is mostly possible to convert results to a different parameterization,
and therefore in principle any theoretician can convert experimental results to another basis,
it is preferable and less error-prone if the experimental result is already expressed in a different parameterization;
publications like that of the ATLAS Collaboration~\cite{Aaboud:2017rwm} go a long way towards the ideal situation.
I reproduce here Table~\ref{tab:oneD_results} and Table~\ref{tab:oneD_EFTLimits}, which express the same experimental results
in the two most common EFT parameterizations~\cite{Baur:2000cx,Degrande:2013kka}.

\newcommand{\fyfour}{\ensuremath{f_4^{\Pphoton}}}
\newcommand{\fzfour}{\ensuremath{f_4^{\PZ}}}
\newcommand{\fyfive}{\ensuremath{f_5^{\Pphoton}}}
\newcommand{\fzfive}{\ensuremath{f_5^{\PZ}}}

\begin{table}[h!]
  \centering
  \scriptsize
\resizebox{0.49\textwidth}{!}{
\begin{tabular}{lll}
\hline
Coupling strength & Exp. 95\% CL $\big[\times 10^{-3}\big]$ & Obs. 95\% CL $\big[\times 10^{-3}\big]$\\
\hline
$f_4^{\gamma}$ & $-2.4$, 2.4 & $-1.8$, 1.8 \\ 
$f_4^{\mathrm{Z}}$ & $-2.1$, 2.1 & $-1.5$, 1.5 \\ 
$f_5^{\gamma}$ & $-2.4$, 2.4 & $-1.8$, 1.8 \\ 
$f_5^{\mathrm{Z}}$ & $-2.0$, 2.0 & $-1.5$, 1.5\\ 
\hline
\end{tabular}
}
\caption{Expected (Exp.) and observed (Obs.) 95\% CL intervals of the aTGC coupling strengths in the parameterization from Ref.~\cite{Baur:2000cx}. Each limit is obtained setting all other aTGC coupling strengths to zero.~\cite{Aaboud:2017rwm}}
\label{tab:oneD_results}
\end{table}

\begin{table}[h!]
  \centering
  \scriptsize
\resizebox{0.49\textwidth}{!}{
\begin{tabular}{lll}
\hline
EFT parameter & Exp. 95\% CL $\big[\mathrm{TeV}^{-4}\big]$ & Obs. 95\% CL $\big[\mathrm{TeV}^{-4}\big]$\\
\hline
$C_{\tilde{B}W}/\Lambda^{4}$ & $-8.1$, 8.1 & $-5.9$ ,  5.9 \\ 
$C_{WW}/\Lambda^{4}$ & $-4.0$, 4.0 & $-3.0$ ,  3.0 \\ 
$C_{BW}/\Lambda^{4}$ & $-4.4$, 4.4 & $-3.3$ ,  3.3 \\ 
$C_{BB}/\Lambda^{4}$ & $-3.7$, 3.7 & $-2.7$ ,  2.8 \\ 
\hline
\end{tabular}
}
\caption{Expected (Exp.) and observed (Obs.) 95\% CL intervals of EFT parameters in the parameterization from Ref.~\cite{Degrande:2013kka}. Each limit is obtained setting all other EFT parameters to zero.~\cite{Aaboud:2017rwm}}
\label{tab:oneD_EFTLimits}
\end{table}

These issues can in my opinion be easily solved by good bibliographic work when preparing documents
for publication, complemented by cross-collaboration agreements in the context of summary plots preparation
as for example the organized efforts by the LHC Top Physics Working Group~\cite{lhctopwg}. 

\section{Estimating parameters: what are we measuring?}

Physicists---particularly those in HEP---regularly call \textit{measurement of a physics observable/quantity} what statisticians call instead \textit{estimate of a parameter}.
While each field is entitled to its quirks, I prefer to stick to the statisticians' terminology.

We aim at connecting the data with a parametric model that might explain them.
If we want to estimate the mass of the Higgs boson by observing its decay into two photons,
we write a model which predicts the number of times we should observe two photons with a certain invariant mass
in the collision events we collect, as a function of a parameter representing the Higgs boson mass.
We can then \textit{fit} the model by, broadly speaking, looking at which of the possible values of
the parameter results in a prediction which is closest to the diphoton mass spectrum we observe in data:
that value is our estimate.
We also want to report an \textit{interval}, that is to quote a range of values for the parameter (usually around the best value)
that are still sufficiently plausible,
representing our uncertainty in the determination of the best value.

The Bayesian school solves this problem by explicitly connecting the probability $P(x|\theta)$ that we observe a given set of data $x$ conditional on the model parameters $\theta$ to assume certain values (\textit{likelihood})
with the probability $P(\theta|x)$ that the model parameters have a certain value conditional on observing a given set of data.
The Frequentist school defines probability through the outcome of repeated experiments; it is also called classical school,
although we could argue that the first probability theory, developed by Laplace, is intrinsically Bayesian \textit{ante litteram}, and that therefore the real classical school would be the Bayesian one~\cite{Sivia}.
Refusing to assign probabilities to the value of a parameter or to a model, the frequentist school ends up considering the likelihood as a mere
function of the parameter for a fixed set of observed data $\mathcal{L}(\theta; x)$, somehow hiding the conditional dependence;
the likelihood is then maximized, and the value of the parameter which maximizes the likelihood $argmax_\theta \mathcal{L}(\theta; x)$
is taken as the estimated value of the parameter.

Lyons~\cite{Lyons:2013ch} argues that the LHC physics program should aim at using both Bayesian and frequentist techniques for estimating parameters and setting intervals,
as doing so would increase the confidence we have in the results; this is easily justifiable by noting that for estimating parameters and intervals the two approaches
usually converge in the limit of infinite data.
Lyons leaves out from this suggestion the problem of testing hypotheses; Cousins~\cite{Cousins:2013hry} indeed shows that hypothesis testing
is the realm where Bayesian and frequentist methods do not in general yield the same answers, in particular not asymptotically.

While defining a common strategy for the discovery of the Higgs boson, the ATLAS and CMS Collaborations~\cite{CMS-NOTE-2011-005} converged on a fully frequentist procedure based on the profile likelihood ratio.
The effect of this agreement are felt even today, when a fully frequentist approach to parameter estimation, interval setting, and hypothesis testing represents the
standard toolbox for computing and reporting most results in HEP. In the next sections I will therefore mostly review the problem of reporting results in connection with frequentist techniques.
D'Agostini~\cite{doi:10.1142/5262} provides an entertainingly critical view of the Bayesian-versus-frequentists dispute in HEP, heavily biased towards Bayesian reasoning.
Trotta~\cite{Trotta:2017wnx} provides a nice review of the use of Bayesian techniques in the astrophysics community, a community that is particularly well-disposed towards these techniques.
Cousins~\cite{Cousins:2018tiz} reviews most statistical methods used in HEP, with profound reflections on several delicate topics.

\section{Uncertainties: show the dirty details}
\label{sec:uncertainties}

In experimental physics, the point estimate of a parameter (a measurement) is meaningless
without an estimate of the uncertainty associated to it.
I prefer to refer to a measurement as the set of a point and interval estimate or as an interval estimate only, when appropriate.

In HEP it is nowadays important to report not only the total uncertainty affecting a given measurement,
but also the individual components that build up the total uncertainty.
This gives hints for further studies---which parts of an analysis could be improved with profit---and highlights
the status of particular topics, included but not limited to the understanding of the current theoretical models.
Perhaps more importantly, reporting the individual components represents a significant step towards broad combinations of the results;
a proper correlation of systematic effects is possible only if the individual components are available.
Combinations are important not only for estimates of the same physical quantity across experiments, but also and perhaps more importantly
for derived quantities (e.g. PDFs or EFT parameters).

Given a set of results it is unfortunately difficult to find consistent reporting of systematic uncertainties.
Too often, we leave the reader with the overall uncertainty on the final result
and with no information about the individual impact on it of each source of uncertainty.
This can be problematic when trying to understand how much a measurement is limited by, say,
a given theory uncertainty rather than another.
Sometimes results are given quoting separately some ``relevant'' sources of uncertainty
(e.g. splitting by {\em experimental}/{\em scale}/{\em PDF}/{\em other}).

We used to compute the total cross section of a given process by applying the na\"ive formula

\begin{equation}
  \label{eq:crossSection}
  \sigma = \frac{N_{data}-N_{bkg}}{\epsilon L}\,.
\end{equation}

The number of signal events is estimated by subtracting the number $N_{bkg}$ of background events from the number $N_{data}$ of observed events
with the measured integrated luminosity $L$.
The acceptance $\epsilon$ accounts for the effect of theoretical branching fractions and of selecting a fiducial region for the measurement.
The fiducial region consists in the generator-level selection which defines the phase space of the measurement;
correcting for the acceptance of this region permits to obtain the total cross section of the process sought,
rather than the cross section corresponding to the fiducial selection only.

Computing a production cross section using Equation~\ref{eq:crossSection} is however becoming a rare occurrence.
We are in the era of profile likelihood ratios, where in an ever-growing number of publications we extract the fiducial cross section
by parameterizing it in a likelihood function which we then fit to the experimental data.
We start from the probability density function (p.d.f.) $p(x|\mu, \theta)$ for the observable $x$ to assume a certain value in a single event,
given a (single- or multi-dimensional) \textit{parameter of interest} (POI) $\mu := \frac{\sigma}{\sigma_{pred}}$ representing a multiplier of the predicted cross section
we want to determine---also called \textit{signal strength}---and a (generally multi-dimensional) \textit{nuisance parameter} $\theta$
representing all the uncertainties affecting the measurement.
We then move to extending it to a data set of many events $X=\{x_1, ..., x_n\}$ by taking the product of the single-event p.d.f.s.
The number of events in the data set is however a random variable itself that follows a Poisson distribution
with mean equal to the number of events $\nu$ we expect from theory; we account for that by using a \textit{marked Poisson model}:

\begin{equation}
  \label{eq:markedPoisson}
  f(X | \nu(\mu, \theta), \mu, \theta) = Pois(n | \nu(\mu, \theta)) \prod_{e=1}^n p(x_e | \mu, \theta)\,.
\end{equation}

Both the parameter of interest and the nuisance parameters act on the individual p.d.f.s for the observable and on the expectation for the number of events in the data set.
Conway~\cite{Conway-PhyStat} reviews the mathematical framework needed for incorporating a systematic uncertainty as a nuisance parameter;
the systematic uncertainty is modelled as an accessory measurement which adds a multiplicative term per each nuisance parameter to Equation~\ref{eq:markedPoisson}.
These terms are constrained in the fit and can be interpreted as a prior probability density that comes from an accessory measurement,
i.e. the nuisance parameter $\theta$ is estimated with an uncertainty $\delta\theta$.
We often assign to the accessory measurement a Gaussian \pdf, unless the nuisance parameter has a physical bound at zero,
in which case we tend to use a log-normal distribution---that rejects negative values of the parameter.

Our likelihood function $\mathcal{L}(\mu, \theta; X)$ consists in Equation~\ref{eq:markedPoisson} regarded as a function of the parameters $\mu$ and $\theta$ for a fixed value of $X$---i.e. for a fixed set of events observed in the experiment.
We define the maximum likelihood estimate for the parameter $\mu$ as $\hat{\mu} := argmax_\mu \mathcal{L}(\mu, \theta; X)$.
This expression still depends on the nuisance parameters $\theta$: we eliminate this dependence by building a \textit{likelihood ratio}

\begin{equation}
  \label{eq:likelihoodRatio}
  \lambda(\mu) := \frac{\mathcal{L}(\mu, \hat{\hat{\theta}})}{\mathcal{L}(\hat{\mu}, \hat{\theta})}\,.
\end{equation}

In this expression the denominator $\mathcal{L}(\hat{\mu}, \hat{\theta})$ is computed for the values of $\mu$ and $\theta$ which jointly maximize the likelihood function.
The numerator is computed as a function of $\mu$ by fixing the parameters $\theta$ to the values $\hat{\hat{\theta}}$ which maximize the likelihood conditional on the fixed value of $\mu$.
The process of eliminating the dependence on the nuisance parameters by taking their conditional maximum likelihood estimate is referred to as \textit{profiling}---hence the name profile likelihood ratio for the method.
The maximum of the likelihood ratio yields the point estimate for $\mu$; the second derivative of the maximum likelihood ratio yields confidence intervals on the parameter $\mu$.
Particular cases---e.g. when the point estimate lies near the physical range allowed for the parameter---are treated with more sophisticated methods.
Cranmer~\cite{Cranmer:2015nia} provides a complete review of the profile likelihood ratio and associated methods.
Bayesians, as reviewed by Demortier~\cite{Demortier:2002ic}, use to eliminate nuisance parameters by marginalization---i.e. integration---rather than by profiling.
The Particle Data Group provides an extensive section on the statistics underlying LHC results~\cite{Tanabashi:2018oca}.

Cross sections can be estimated in different final states and then combined into a single estimate for increased precision.
Computing a cross section by using the na\"ive Equation~\ref{eq:crossSection} or by doing a profile likelihood fit can yield some residual discrepancy when extracting the uncertainties.
While most analyses nowadays use profile likelihood fits, the 13 TeV determination of fiducial cross sections for associated WZ production
represents an interesting example of methodological discrepancy.
The ATLAS Collaboration~\cite{Aaboud:2019gxl} quotes symmetric uncertainties computed using HERA-era methods
by propagation of the individual uncertainties in each of the parameters entering Eq.~\ref{eq:crossSection}, 
whereas the CMS Collaboration~\cite{Sirunyan2019} opts for asymmetric intervals
from a profile likelihood fit in each channel, and a simultaneous fit to the four analysis channels to obtain the combined result.
How well symmetric uncertainties describe the situation depends on how realistic the underlying assumption is that the likelihood function has a
Gaussian form---or that the profile likelihood as a $\chi^2(N_{POI})$ form, as we will see in Section~\ref{sec:searches}.

Sometimes we estimate simultaneously the POI and a parameter that normally would be classified as a nuisance parameter,
as the CMS Collaboration~\cite{Chatrchyan:2012cz} does by measuring simultaneously the top quark mass and the jet energy scale.
Even when the analysis is not sensitive enough to treat an uncertainty as POI and measure it,
the parameter can still be constrained while profiled in the fit,
indirectly giving information---when appropriate---about the amount of over- or under-estimation
that went into the determination of its value prior to the fit ({\em pre-fit} value, as opposed to {\em post-fit} value,
that corresponds to the best fit value for a given profiled nuisance parameter as determined by the fit procedure).
This in turn can inform theoretical developments in the modelling of some sources of uncertainty
(PDF, $\alpha_{s}$, ISR/FSR, heavy flavours).

It is therefore crucial to report in detail both the pre- and the post-fit values of the uncertainties,
split by uncertainty source. Many analyses tend to quote only the total uncertainty on the final result;
this is understandable historically, because we often propagate in bulk the uncertainties in the efficiencies $\epsilon$ of Eq.~\ref{eq:crossSection}
to the final measurement and the full breakdown of the sources of systematic uncertainty
is not available---unless we perform the combination of the results
with methods like BLUE~\cite{aitken_1936,LYONS1988110,Lista:2016vxe}.

Barlow and colleagues~\cite{babar} argued in a document of the BaBar Collaboration that split uncertainties should be quoted only when an important source of uncertainty
is poorly controlled and will be likely updated after publication;
this would allow for an easy recomputation of an updated uncertainty in the published result.
When the result originates from a maximum likelihood fit, however, the correlation scheme among systematic uncertainties after the fit implies that
simply replacing the original uncertainty with a recomputed one is a questionable procedure;
it is better to reperform the whole analysis with updated inputs to reflect the new knowledge of the uncertainty source.
Cranmer and colleagues~\cite{Cranmer:2013hia} advocate using a technique they called \textit{recoupling},
where the likelihood is parameterized as a function of some theory effect (e.g. particles decay rates);
the publication of the parametric likelihood permits the recomputation of the result with improved theoretical modelling
that might be available years after the experimental analysis is published.

The already cited BaBar document on statistical practices also recommends combining results affected by asymmetric systematic uncertainties
by adding in quadrature the uncertainties separately for each side. Although they warn that this is a non-theoretically-justified convention
and that sometimes the non-gaussianity of the likelihood might be too pronounced for this procedure to make sense, they state that
\textit{there is no better alternative and it is a convention}.
While this might have been true a few decades ago, nowadays it is very easy to combine measurements by merging the
individual likelihoods into a combined one which we then refit to obtain the combined measurement and well-defined uncertainties:
computationally, we have vast available resources to make fits with thousands of nuisance parameters;
practically, we have accumulated vast experience in structuring our likelihoods in ways that ease the task of combining them with those of other experiments.
I argue therefore that it is crucial that the HEP Collaborations publish the full likelihood for all their measurements.
The ATLAS Collaboration has started publishing the full likelihood of certain measurements, and provided a concrete
example of reproducibility enabled by the publication of the full information~\cite{ATL-PHYS-PUB-2019-029}.

The \textsc{HepData} project~\cite{Maguire:2017ypu} facilitates sharing this kind of information;
the format already supports matrix representations (useful for displaying correlation among uncertainty sources),
and the \textsc{Rivet} project supports already the breakdown of uncertainty sources via the \textsc{Yoda} format~\cite{yoda}.
While the tools are available already, a strong push is still needed across the Collaborations to publish a larger-than-minimal set of information.
We should always quote the pre-fit uncertainties in the publication text,
together with explicit mention to the way we computed them.
We should similarly always quote post-fit uncertainties, I suggest in a tabular way, and split them into as many groups
of independent components as it makes sense to quote separately.
The CMS Collaboration does this e.g. in Ref.~\cite{Sirunyan2019};
this is crucial for understanding the intimate characteristics of a result, and provides hints for future developments.
This splitting gives an overview of how much of the total uncertainty is imputable to a given set of systematic effects:
we freeze these uncertainties to their post-fit values, repeat the fit to extract a new (smaller) uncertainty,
and obtain the contribution to the overall uncertainty as a squared difference between the full and reduced
uncertainties---we obtain the contribution of the statistical uncertainty by freezing all the nuisance parameters.
This is the modern expression of what Fisher called \textit{``the constituent causes fractions or percentages of the total variance which they together produce''}~\cite{fisher_1919}
and \textit{``the variance contributed by each term, and by which the residual variance is reduced when that term is removed''}~\cite{fisher_1921}
when formalizing and applying the concept of variance to develop the analysis-of-variance (ANOVA) method.
The best way of reporting post-fit uncertainties is to quote also what we call the \textit{pulls} and \textit{constraints}
of the nuisance parameters in the statistical model, and their \textit{impacts} on the post-fit signal strength.
We define the (normalized) \textit{pull} of a nuisance parameter as the difference of the post-fit and pre-fit values
of the parameter, normalized to the pre-fit uncertainty, $pull := \frac{\hat{\theta}-\theta}{\delta\theta}$,
and the \textit{constraint} as the ratio between the post-fit and the pre-fit uncertainty in the nuisance parameter.
We can therefore easily spot possible issues in the fit: a nuisance parameter that is pulled too much
may be a hint that our estimate of the pre-fit value was not reasonable; a nuisance parameter
that is constrained too much indicates that the data contain enough information to improve the precision
in the nuisance parameter with respect to our original estimate, which may or may not make sense.
The \textit{impact} of a nuisance parameter on the post-fit signal strength permits to obtain a ranking
of the nuisance parameters in terms of their effect on the signal strength;
we fix each nuisance parameter to its post-fit value $\hat{\theta}$ plus/minus its pre-fit (post-fit) uncertainty $\delta\theta$ ($\delta\hat{\theta}$),
reperform the fit, and compute the impact as the difference between the original fitted signal strength
and the refitted signal strength.

\begin{figure}[!t]
  \centering
  \includegraphics[width=\linewidth]{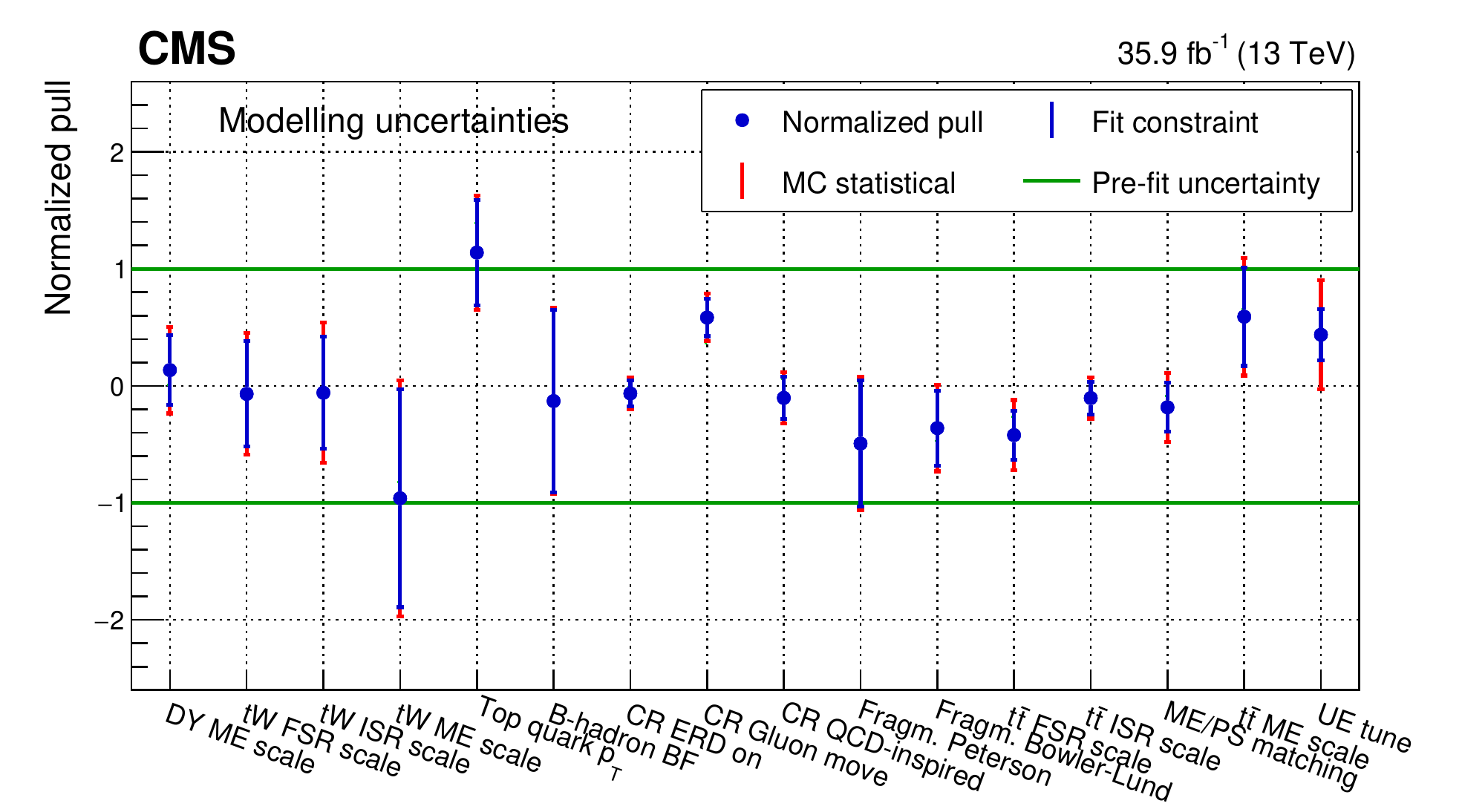}\\
  \caption{The normalized pulls and constraints for a simultaneous fit of the top-antitop production cross section and top mass~\cite{Sirunyan:2018goh}.
    The 
    The leading lepton transverse momentum distribution in a validation region (top) and the dijet mass distribution in a signal region (bottom), in a publication reporting evidence for triboson production in ATLAS~\cite{Aad:2019udh} (v1). In the top figure the hatched \textit{Uncertainty} is statistical only, whereas in the bottom figure it is the squared sum of the statistical and systematic uncertainties. Later versions of the publication (v3) now explicitly mention \textit{Stat. uncertainty} in the legend of the top plot.}
  \label{fig:constraints}
\end{figure}

I still observe some resistance to showing the full set of pulls, constraints, and impacts, due to the worry that reporting the pulls would somehow make the result look less solid.
On the contrary, a result is more solid if it is accompanied by a complete analysis of the uncertainties that affect it.
Nevertheless, recently an increasing number of publications started to contain the full information.
The CMS Collaboration published the pulls and constraints (Figure~\ref{fig:constraints}) for a top-antitop pair production cross section
fit in events with two leptons~\cite{Sirunyan:2018goh};
this is an example where the severe constraints on the nuisance parameters related to the top-antitop signal
express \textit{``the strength of the analysis ansatz''}~\cite{Sirunyan:2018goh}.

\begin{figure}[!t]
  \centering
  \includegraphics[width=\linewidth]{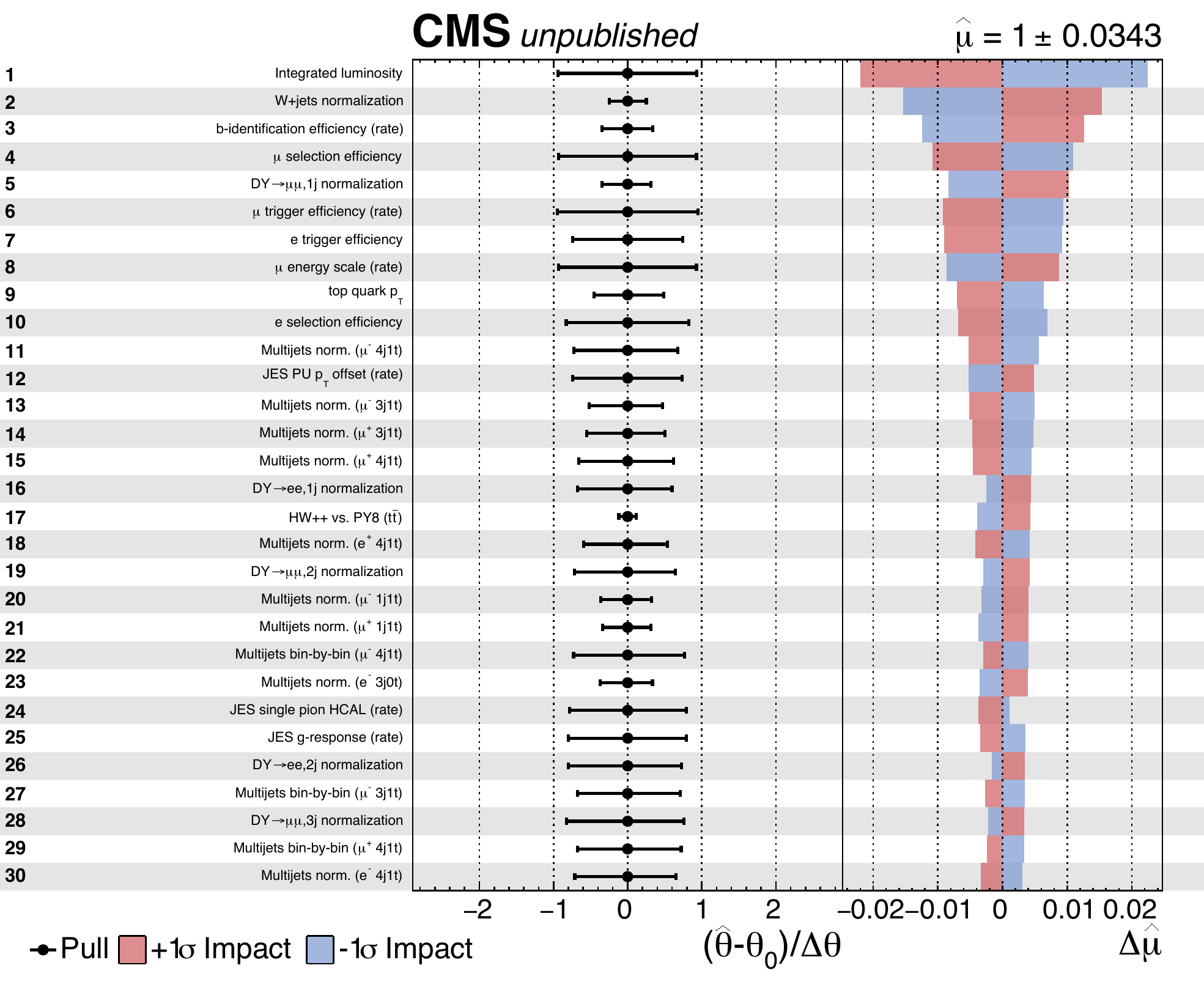}\\
  \caption{The so-called \textit{impact plot}, which reports the pulls, constraints, and impacts of the nuisance parameters for a fit to the Asimov data set for a top-antitop cross section measurement in events with one lepton in the final state~\cite{Sirunyan:2017uhy}.}
  \label{fig:asimovimpacts}
\end{figure}

The CMS Collaboration also published the full information (pulls, constraints, and impacts)
measuring the top-antitop production cross section in events with one lepton in the final state~\cite{Sirunyan:2017uhy}.
Although the impacts are published only as \textit{additional material} in a dedicated webpage~\cite{pubpagetop-17-006},
this example is particularly commendable because the CMS Collaboration published not only the
observed impacts (which provide indication on which constraints are induced by the observed data),
but also the impacts (Figure~\ref{fig:asimovimpacts}) on the \textit{Asimov data set}---a data set we obtain by replacing the data with the expectations from simulated events:
a fit to the Asimov data set should yield a ``perfect'' (zero pulls, zero constraints) result by construction.

\begin{figure}[!t]
  \centering
  \includegraphics[width=\linewidth]{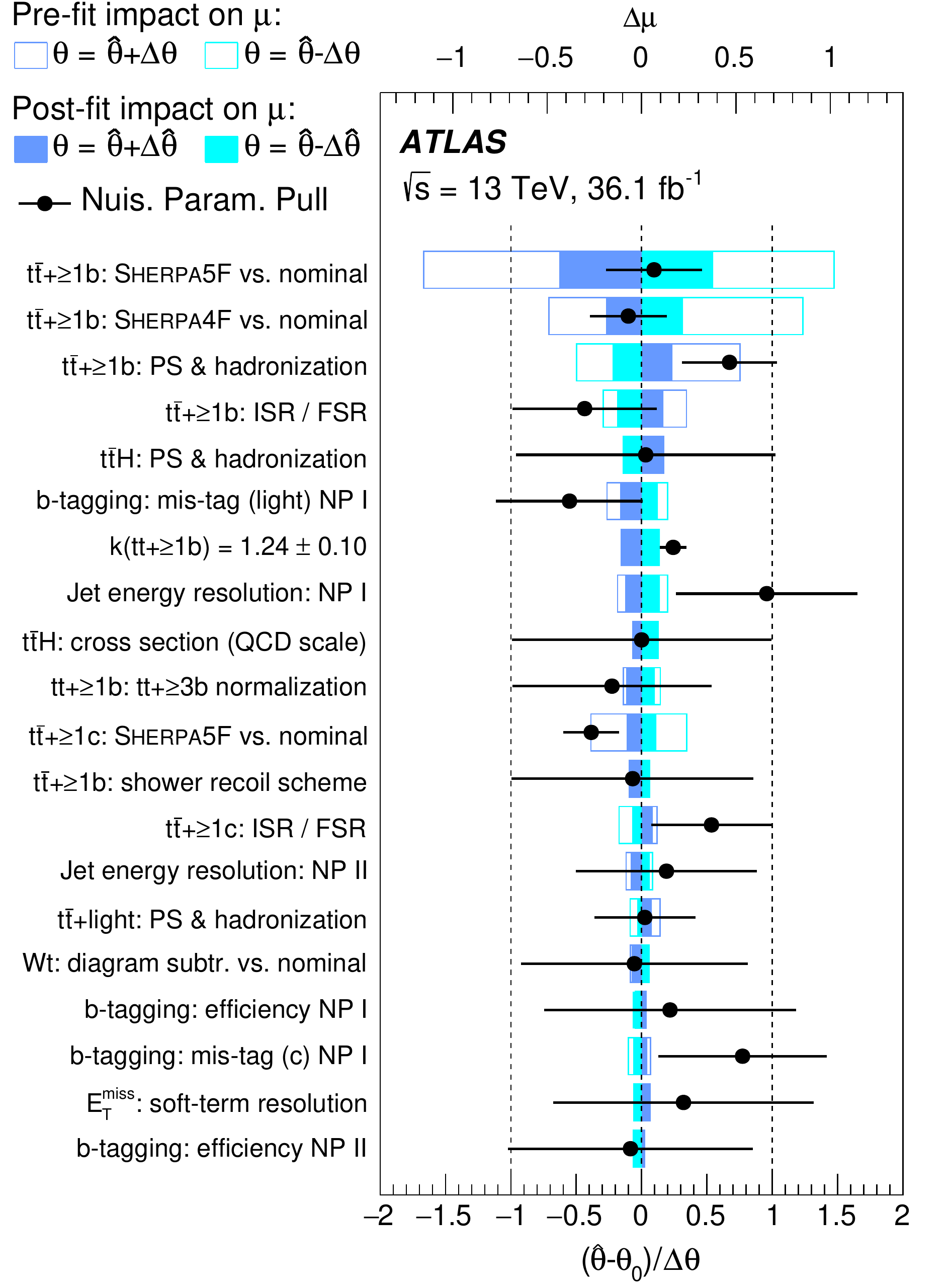}\\
  \caption{The impact plot, reporting pulls, constraints, and impacts of the nuisance parameter for a fit of the top-antitop-Higgs production cross section for a Higgs boson decaying into two bottom quarks.~\cite{Aaboud:2017rss}. The four most important nuisance parameters gave rise to a deep debate on the modelling of the main background of the analysis.}
  \label{fig:ttHbb}
\end{figure}

The ATLAS Collaboration published impact plots when searching for associated top-antitop-Higgs production (ttH) when the Higgs boson decays to two bottom quarks~\cite{Aaboud:2017rss}.
The selected decay of the Higgs boson gives rise to a topology which is very similar to that of a top-antitop pair when additional bottom quarks are produced from the gluons initiating the process;
the systematic uncertainties in the production of additional bottom quarks in top-antitop events are dominant. Figure~\ref{fig:ttHbb} gave rise to an interesting debate on the modelling of
the main background at the 2018 Higgs Toppings Workshop~\cite{higgstoppings} in the lovely Benasque, highlighting the importance of publishing the full information to fuel theoretical and experimental debate within the HEP community.

Including a theory uncertainty as a nuisance parameter in the fit is sometimes a suboptimal choice.
A prototype case is that of the \textit{two-point systematic}:
you may be aware of a possibile mismodelling, but your systematic variation might just consist
in a different generator model.
The aforementioned case of the ttH (H$\to$bb) ATLAS search~\cite{Aaboud:2017rss} exhibits as most important nuisance parameters
those corresponding to the two-point systematic consisting in the difference between the nominal background templates
obtained with the \textsc{POWHEG} generator and alternative templates obtained with the \textsc{SHERPA} generator.
The concept of \textit{variation} for that uncertainty is ill-defined, and modelling the uncertainty as a nuisance parameter
with a certain prior does not really capture the problem; assuming a Gaussian shape for the prior density cannot be justified,
and arguably a continuous transformation of the nominal template into the alternative one might even not exist.
While it is true that a continuous transformation can be almost always found for example via optimal transport metrics like the
Wasserstein distance~\cite{7974883}, the physical meaning of such a transformation is problematic; it is not guaranteed
that any alternative generator would lie along the transformation between the two reference ones, and it is not even
guaranteed that the underlying true shape from nature would lie along that ransformation.
I argue that in this cases it is best to \textit{externalize} the problematic nuisance parameters.
This procedure consists in removing the nuisance parameter from the statistical model, and performing two separate fits:
in the first fit, we model the relevant process using the nominal template; in the second fit, we model the relevant process using the alternative template.
The difference between the two fitted signal strengths gives an estimate of the systematic uncertainty in the process modelling.
In the more dramatic cases I think that the two separate results should be quoted,
but most of the times quoting the difference as a systematic uncertainty assigned to the result of the nominal fit is fine.
The CMS Collaboration resorts to externalization in one of its measurements of the single top quark production in association with a W boson~\cite{CMS-PAS-TOP-19-003}.

Sometimes the post-fit uncertainties are expressed in terms of their effect on the yields
rather than on the POI, as the ATLAS Collaboration did~\cite{Aaboud:2017rwm}.
However, while it is interesting to see a table of the post-fit yields
and their total uncertainty, in the era of differential measurements and measurements of the couplings
I find more useful to quote the impact on the POI rather than on the yields.

The individual pulls and impacts, while useful, do not however represent the full picture
because they do not inform about the correlation among the various sources of uncertainty;
I therefore further advocate for the publication of the full post-fit covariance matrix, split by uncertainty sources.
Because of these correlations, when we publish impact plots we should also make explicit that the uncertainties
cannot be summed in quadrature: the resulting estimate of the total uncertainty would neglect correlation effects
and would be often significantly different from the correct estimate from the curvature of the full profile likelihood.

The CMS Collaboration~\cite{Collaboration:2242860} has proposed a simplified-likelihood approach that has been improved by Buckley and colleagues~\cite{Buckley:2018vdr}.
They collapse all the systematic uncertainties into individual per-search-region variations of the yields
and write a simplified likelihood that accounts for the correlations of the yield variation across the regions;
they then merge the search regions into aggregated regions in order to simplify reinterpretations.
While this approach maintains the information on the correlation between regions
(although losing some of it when aggregating regions), it gains in computation efficiency
and in simplicity of the published modelling at the price of losing all information relative to the individual contribution
of each source of uncertainty to the final measurement.
This is certainly a better approach with respect to not publishing even the correlation of the yields among search regions,
but it does not necessarily provide all the information needed to perform a combination with other results;
it nevertheless provides a simple way for third parties to test the sensitivity of new models
in an approximate way
without having to simulate the full detector, having therefore a high practical value.
These simplifications can however be made starting from the full published covariance matrix
with all the separated sources of uncertainty,
and I therefore regard this method as a practical addendum to the disclosure of the full information. 

\section{Searching, often not finding}
\label{sec:searches}

HEP scientists not only perform precision measurements for phenomena predicted by the SM,
but also search for yet-to-be-observed phenomena beyond those predicted by the SM.
For a measurement they usually report a point estimate and a two-sided confidence interval.
For a search they want to usually report an upper (or lower) limit, that is a one-sided confidence interval
whose other side is the physical boundary for the parameter we seek to estimate.

The signal strength $\mu:=\frac{\sigma}{\sigma_{SM}}$ discussed in Section~\ref{sec:uncertainties} has a physical boundary at $0$,
because a cross section (ratio) is definite positive.
Although from the point of view of frequentist statistics there is no conceptual issue in reporting a negative estimate for a positive-definite parameter
(it corresponds to sampling from the \pdf of possible data), 
it is preferred in these cases to quote an upper limit bounded at zero.
The usual situation is that a signal is not visible with a certain data set, and we have to quote an upper limit to its production cross section.
We later improve the experiment (e.g. by collecting more data or by reducing the main systematic uncertainties), until
the moment when we start ``seeing'' an excess of events which at some point will hopefully become so large that
we can just assume the signal is there.
We need therefore a rule for deciding when do we keep quoting an upper limit and when we start quoting a point estimate and a two-sided interval.

The smooth transition from upper limits to two-sided intervals is a problem solved by the Neyman construction~\cite{10.2307/91337} and the Feldman-Cousins ordering principle~\cite{Feldman:1997qc} for general cases.
In the case of a search for new physics, for example in the case of the Higgs boson, we are interested in something more.
On one side we want to maximize our exclusion potential when searching for a non-existing signal in a region in which we would be sensitive to it;
on the other side we want to maximize our discovery potential when searching for an existing signal in a region in which we would be sensitive to it.
This problem has been mostly solved by Read~\cite{Read:451614,Read_2002} in the context of Higgs searches at LEP;
the $CL_s$ algorithm rescales the probability of a fluctuation under the signal-plus-background model by the
probability of a fluctuation under the background-only model.

Finally, we assess whether we can start claiming the observation of a signal by quoting its \textit{significance}, which is the probability
under the hypothesis of no signal that the data fluctuate as much or more than observed.

The $CL_s$ method, as well as the computation of the significance, relies on knowing the distribution of a suitable test statistic (normally the likelihood ratio) for the null hypothesis;
we build the distribution by running often computationally-intensive toy experiments, a procedure we refer to as \textit{full $CL_s$} method.
If the number of observed events is high enough, Cowan and colleagues~\cite{Cowan:2010js} have shown that by the Wilks' theorem the distribution of the test statistic asymptotically approaches
a $\chi^2(N_{POI})$.
When the Wilks' theorem is valid, we can therefore compute tail areas to a good approximation by integrating the $\chi^2$ distribution
from the observed value of the test statistic to infinity, rather than generating computing-costly toy experiments.
The goodness of the approximation should be assessed by comparing the asymptotic $CL_s$ results (median expected limits, for example)
to full $CL_s$ results in at least the most delicate cases. This comparison is particularly necessary if the bulk of the sensitivity
of the analysis comes from regions with a low number of events.

The asymptotic $CL_s$ method works surprisingly well even for a relatively low number of events, yielding in most cases differences of a few percent with respect to the full construction.
In some dramatic cases however, the asymptotic construction can yield overoptimistic limits by about $20$--$30\%$: how to deal with this issue?
When the sought signal is very well defined---as is the case e.g. for most SM or Exotica Higgs analyses---Collaborations tend to quote asymptotic $CL_s$ limits only when
the difference with full $CL_s$ is $\mathcal{O}(1\%)$, and to compute the full $CL_s$ for all the hypotheses if the discrepancy is larger.
In some class of analyses we do not know the signal model to a great precision.
Many searches for supersymmetry (SUSY) use for example Simplified Model Spectra~\cite{Alwall:2008ag,Alves:2011wf}, where
predictions are computed assuming that only one SUSY particle exists in each simplified model: the presence of other SUSY particles predicted by the full model is neglected.
These analyses are also characterized by a huge number of alternative (signal) models, and computing the full $CL_s$ construction for each of the models
is sometimes computationally difficult.
In such cases we still compare the full and asymptotic $CL_s$ constructions in a handful of regions where we expect the worst discrepancies,
and quote the asymptotic results everywhere if in the tested regions
they do not differ from the full construction by more than about $20$--$30\%$;
we assume that the simplifications involved in building the simplified model spectra are larger than the discrepancy between full and asymptotic $CL_s$.

Theory predictions for a given new physics signal model normally depend on one or more theory parameters.
We often compare the obtained observed upper limit with the theory prediction to determine which is the parameters range
where the theory prediction is higher than the observed limit; that range is the region where we exclude the signal model at a given confidence level (nowadays usually 95\%, a few decades ago usually 90\%).
When the theory prediction is affected by uncertainties on its calculation,
we need to take into account this uncertainty when defining the exclusion region.

The best approach is to incorporate the theory uncertainty as a nuisance parameter;
if the parameter is either integrated out in a Bayesian construction or profiled in a likelihood ratio fit,
the result of the procedure will already include the effect of this uncertainty,
and the scientist can simply compare the limit with the nominal prediction.

If the uncertainty is too large, integrating or profiling the parameter (which can both be seen as forms of averaging it)
are not really representative of the real effect of the uncertainty. In a sense, when the uncertainty is too large
then it typically makes more sense to consider two (or multiple) alternative models and report separate results;
I described this externalization approach towards the end of Section~\ref{sec:uncertainties}.

A similar situation is recommendable in case the theory model is so simplified---as is the case of te Simplified Model Spectra described
above---that the theory uncertainty may not be representative of the real systematic effect.

\begin{figure}[!t]
  \centering
  \includegraphics[width=\linewidth]{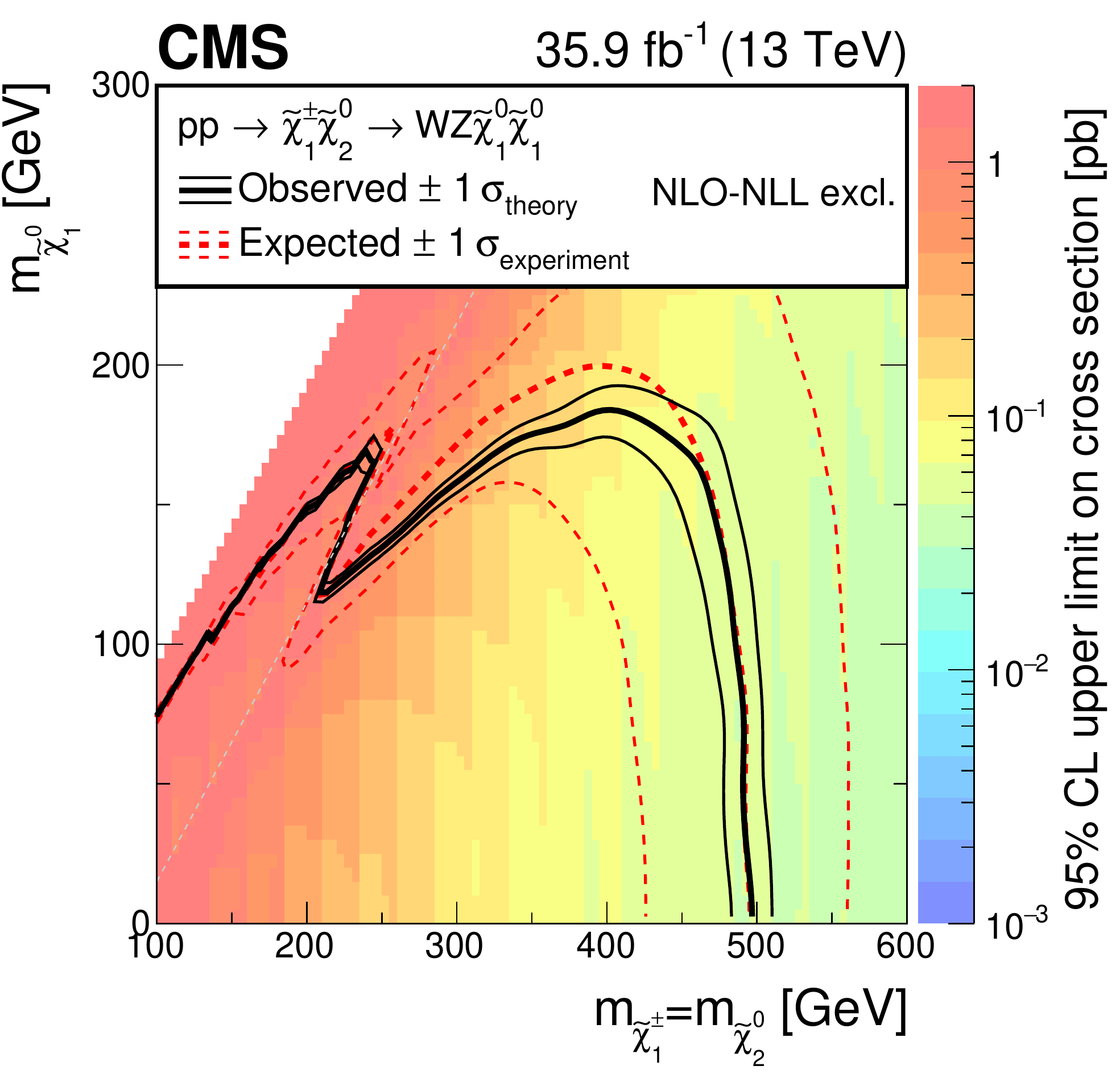}\\
  \caption{The 95\% confidence level upper limit on the production cross section of a Supersymmetry signal~\cite{Sirunyan:2018ubx}. The observed limit is complemented with a theory uncertainty band (in black), while the expected limit is complemented with an experimental uncertainty band (in red).}
  \label{fig:susyewkino}
\end{figure}

SUSY analyses in the ATLAS and CMS Collaborations tend to incorporate the uncertainty on the theory cross section by computing expected
and observed limits without theory uncertainties in the signal model, and adding theory uncertainty bands around the observed limit
and experimental uncertainty bands around the expected limit, as Figure~\ref{fig:susyewkino} illustrates~\cite{Sirunyan:2018ubx}.
It is not straightforward to disentangle when such a procedure is well-founded according to the principles above and when it is a mere convenient choice.

\section{Phase space: why is it even different?}

The extrapolation of fiducial cross section measurements to the full phase space depends crucially on the definition
of the fiducial region. When comparing confidence intervals on EFT parameters with theory predictions,
as in the \textsc{SFitter} package~\cite{Biekotter:2018rhp,Biekotter:2018jzu},
it is likewise necessary to have access to the exact definition of the fiducial region used to compute
the appropriate confidence intervals.

ATLAS and CMS most of the times use---for a given analysis---different definitions of the phase space.
Some are dictated by the different characteristics of the detectors,
but some other could be uniformized across the experiments.
In the latest WZ measurement the ATLAS Collaboration~\cite{Aaboud:2019gxl} selects a Z boson mass interval of 66--116 GeV,
whereas the corresponding interval for the CMS Collaboration~\cite{Sirunyan2019} is 60--120 GeV;
there is a feedback loop in which theoreticians like Grazzini and colleagues~\cite{Grazzini:2017ckn} are forced to make predictions for two different definitions
coming from the experiments,
and the experiments end up perpetuating the discrepancy by publishing new results
using a phase space corresponding to the available predictions.

Any improvement in this direction would simplify the work of theoreticians,
both in producing and comparing predictions for the different phase spaces
and in reproducing the fiducial cuts when importing experimental results into global fits.
Some sub-communities exhibit already a quite high degree of synchronization in terms of phase space,
but for others more effort is needed. The ATLAS Collaboration~\cite{Aaboud:2017qkn} sets an exemplary standard for reporting
phase space definitions, as Table~\ref{tab:fiducialtruthlevelcuts} illustrates.

\begin{table}[!htbp]
\begin{center}
\resizebox{0.49\textwidth}{!}{
\begin{tabular}{c|c}
\hline\hline
 Fiducial selection requirement  & Cut value \\
\hline\hline
 $p_T^{\ell}$    &       $>25$ GeV  \\
$ |\eta_{\ell}|$                                        &       $<2.5$                  \\
$ m_{e\mu}$                             &       $>10$ GeV               \\
$N_{jets}$ with $p_T>$ 25(30) GeV, $|\eta| <$ 2.5(4.5) & 0 \\
$E_{\mathrm{T,\,Rel}}^{\mathrm{miss}}$                                               &       $>15$ GeV \\
$E_{\mathrm{T}}^{\mathrm{miss}}$                                          &       $>20$ GeV \\
\hline\hline
\end{tabular}
}
 \end{center}
\caption{Definition of the $WW\rightarrow e\mu$ fiducial phase space, where $\ell=e$ or $\mu$~\cite{Aaboud:2017qkn}. }
\label{tab:fiducialtruthlevelcuts}
\end{table}
An especially good development is represented by the current push to publish the details
of phase space definitions as a routine in the \textsc{Rivet} software~\cite{Buckley:2010ar}.
CERN Yellow Reports like the ones produced by the LHC Electroweak Working Group~\cite{lhcewwg} are another obvious place to document and cite such agreements---provided that
later analyses apply these recommendations.

\section{Machine Learning: do not let it be a black box}

The proliferation of the use of machine learning methods
in HEP and their successes have resulted in a desensitization of the community to the internal details of its tools.

While this is a testimony to the trust the community puts into methods that 10--15 years ago
were looked at with suspicion by a large part of it, on the other side this implies that entire parts
of the employed analyses are described only very succinctly in the publications.
In case of deep artificial neural network (\textit{``deep learning''}), their relative novelty in HEP
fuels the release of material with a certain level of detail; documents tend to cite publications
that describe a very specific flavour of neural network, which helps the reader
to get familiar with a specific technique rather than the broad concept of deep learning.
In these networks neurons are layered, and different flavours of network differ typically by structure of the neuron connections;
when neurons in a layer are not connected to all the neurons in the following layer,
the network becomes specialized at certain tasks (image recognition, temporal sequence recognition, etc) that depend on the configuration of the connection.
Goodfellow and colleagues~\cite{Goodfellow-et-al-2016} provide a sufficiently recent review of the field. 

In case of simpler classifiers like boosted decision trees (BDTs), instead,
the level of detail nowadays often does not go beyond stating that a BDT method has been applied to solve the problem at hand.
Even analyses in which BDT classifiers are crucial to reach a given
sensitivity threshold---as that in Ref.~\cite{Aad:2019udh}---sometimes contain no information
either on the gain given by using the classifiers or on the details of the structure and methods used to train them.

I propose that at the very minimum some details on the training procedure are provided in the publication text
for each classifier used in the analysis, answering to these questions: ``Are generators used for training and validation'',
``When the same generator is used, how are the events split into training, validation, and application data sets'', and
``when different selections are used for training/validation than for application, how are extrapolation effects accounted for''.
An example of a reasonable amount of clear information is Ref.~\cite{Sirunyan:2019arl}.

In case training and application data come from different sources,
a discussion of the possible effects on the relevant quantities should also be included.
The way the relevant hyperparameters of the classifier have been chosen (Bayesian optimization, grid searches, etc.)
should be definitely quoted.
All of this is interesting information per-se and can (and should!) also inform further needs for future improvements
of such methods. Validation of the classifiers---although possibly only for classifiers used for later statistical
inferences---should also be highlighted; the agreement between data and simulation should be studied
in control regions and shown in the signal region.

Publishing the trained model or set of training weights can probably be counterproductive;
a legitimate worry is that classifiers could be used improperly by colleagues trying to interpret the results
in a non-suitable way or even apply the classifiers at particle level or with some different simplified detector simulations. This is a point that
should be somehow discussed, however, particularly in an era in which any statistical learning
development is published together with any possible detail; would a disclaimer about the usage of
such information be enough to discourage improper use?
The authors of manuscritps on machine learning methods often nowadays publish a fully reproducible workflow (including the input data) as companion software, as the authors of \textsc{INFERNO}~\cite{deCastro:2018mgh} and those of \textsc{CWoLa}~\cite{Collins:2019jip} do.
Some machine learning techniques are applied to cases that are very specific to the experimental environment,
and a handy way of applying these techniques to realistic cases without having to publish within a Collaboration consists in using the so-called \textit{open data}; experimental Collaborations tend to publish old data sets---after having squeezed them to the last bit for discoveries and measurements---or specific data sets specifically for machine learning purposes.
Alison and colleagues~\cite{Alison:2019kud} have developed and algorithm aiming at an end-to-end particle and event identification (that is reconstructing the full topology of a collision in the detector by using an individual adaptive algorithm rather than reconstructing each type of particle in stages), and they applied it to Open Data from the CMS Collaboration (incidentally, without publishing a reproducible workflow).
When instead members of a Collaboration develop or apply dedicated machine learning techniques
to deploy them within the experiments's reconstruction software,
generally the result is a publication signed by the whole Collaboration as in Ref.~\cite{Sirunyan:2017ezt}
and neither the data nor the algorithms are public.
If the algorithms are related to the data taking (as is the case for trigger systems and object reconstruction) they often end up to be part of the sometimes public Collaboration's code~\cite{thecmssw},
but this situation is normally not referenced explicitly.

\section{Unfolding: if you do it, try not doing it wrong}

Unfolding is an inverse problem that sometimes should not be solved in fundamental physics research,
as it always comes with a certain amount of arbitrariness in its application.
In HEP, however, we often have compelling practical reasons for unfolding spectra,
mostly gravitating around enabling future consumption of the results.
Any result obtained using the particles reconstructed in the detector will depend on the details of the detector response to the particles produced in the collisions,
because a particle will be reconstructed in the detector with kinematic properties different than its original ones.
When looking at a differential distribution in some observable, the presence of the detector will therefore induce migration effects
(a particle with a transverse momentum lying in a given bin of the distribution will be reconstructed with a different momentum which may
lie in a different bin).
We model the detector response as a \textit{response matrix} defining the probabilities of migration from the \textit{generator-level} bins to the \textit{reconstruction-level} ones; we are interested into the inverse mapping, to convert reconstruction-level distributions into generator-level ones, and therefore we need to invert the response matrix.

The simplest example of response matrix inversion is the analytic inversion of a $2\times2$ matrix; the solution depends on the individual terms
of the matrix as well as on the reciprocal of the discriminant:

\begin{equation}
  \begin{bmatrix}
    a & b \\
    c & d \\
  \end{bmatrix}^{\!-1} = \frac{1}{ad-bc}\begin{bmatrix}
    d & -b \\
    -c & a \\
  \end{bmatrix}\,.
\end{equation}

\begin{figure}[!t]
  \centering\includegraphics[width=0.8\linewidth]{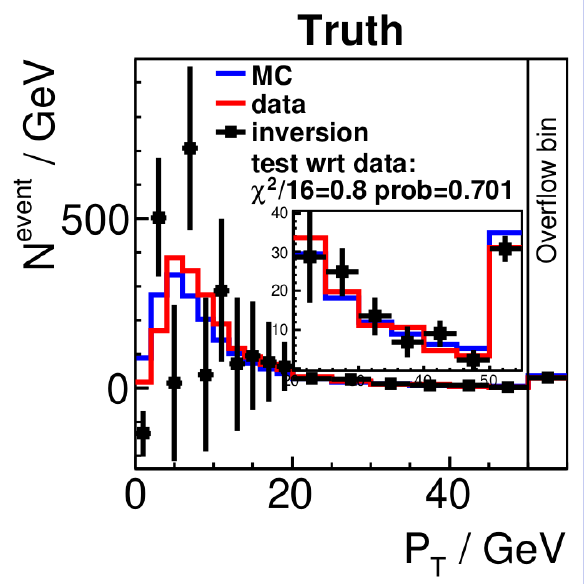}\\
  \caption{Unfolding by matrix inversion~\cite{Schmitt:2016orm}. A small determinant results in large fluctuation of the unfolded result (black points), illustrating the need of regularization techniques.}
  \label{fig:unfoldoscill}
\end{figure}

If the determinant is close to zero, its reciprocal may be very large and the reconstructed distribution could be mapped into a widly fluctuating unfolded distribution. This problem is present also at higher dimensions (Figure~\ref{fig:unfoldoscill}).
Matrix inversion yields an unbiased solution and is therefore preferrable: if the response matrix is not (almost) diagonal,
large fluctuations might appear.

\textit{Regularization} is an artificial constraint introduced to fix these high-frequency fluctuations;
the variance of the unfolded solution is kept at bay, but the procedure introduces a bias towards the simulated distribution.
This bias can be studied, but regularization should nevertheless be used with care and only when necessary.
For many cases in which the response matrix is almost diagonal, regularization does not improve the measurement.

Likelihood-fit-based unfolding embeds regularization as an explicit Tikhonov term~\cite{tikhonov} whose strength can be tuned
by tweaking a parameter called \textit{regularization strength}---this method is commonly known in statistics literature
as ridge regression~\cite{doi:10.1080/00401706.1970.10488634}.
The value of the regularization strength is chosen by scanning some quantity sometimes related to a $\chi^{2}$.
Other algorithms are based on iterating to convergence; regularization is introduced in the form of early-stopping,
as for example is the case for the D'Agostini method~\cite{DAgostini:1994fjx},
where iteration is deemed necessary to reduce effects from the dependence of the posterior on the suggested flat prior.
Its improved version~\cite{DAgostini:2010hil} consists in careful modelling of some effects
which were approximated with Gaussian distributions in the original version of the algorithm,
and suggests that a careful modelling of the prior fed to the first iteration of the algorithm (flat in the original version)
might help in avoiding or (in practice) at least reducing the need of iterations.

D'Agostini~\cite{DAgostini:2010hil} makes an important point while describing the eponimous algorithm: \textit{``My recommendation is that one should know what one is doing''}.
Schmitt~\cite{Schmitt:2016orm} has indeed highlighted that iterative methods converge in a number of steps that depends on the problem:
it is therefore important to mention both the value used for the parameter describing the number of iterations,
and most crucially the stopping criterion used to determine that parameter.
It is not infrequent in HEP literature that the software default (4, for the standard D'Agostini implementations) is used,
but that value does not necessarily represent the best choice for the problem at hand;
sometimes more iterations are needed to achieve convergence, sometimes even less.
For some algorithms, \cite{Schmitt:2016orm} points out, increasing too much the number of iterations can even worsen the result.
Not choosing correctly the number of iterations can induce unwanted and non-studied regularization biases,
and failing to report the number of iterations and the method used to determine it might make the result untrustable. 

Agreeing on the details of the procedures is an open topic, discussed often in the ATLAS and CMS statistics communities;
mostly we have prescriptions on what \textit{not} to do rather than unique recommended ways of doing something
(unfortunately references are password-protected).
Leaving aside the open questions on the procedure itself, when it comes to reporting on it
we should nevertheless come together as a community on a minimal set of details to be reported for any unfolding measurement.

We should first of all always publish the response and covariance matrices.
It is true that in many cases the response matrix is available as supplementary material in \textsc{HepData}, and that
some journals give the possibility of submitting additional material with its own DOI code,
but while this is a step in the good direction I argue that both response and covariance matrices
should have their space in the publication body itself (maybe as an appendix), as they are critical to understanding the application
of the procedure to the data at hand.

Results should be published not only in graphical form, but also in tabulated form;
the recent push of the Collaborations to fill the \textsc{HepData} entries for each new result
is definitely the symptom of a more attentive attitude towards the theory community.

I finally suggest that the minimal amount of information to be published on any unfolding procedure
should include also a description of the settings used for any given method;
the choice of introducing a regularization parameter should be first of all justified---regularization
should be used as originally intended as a way of dealing with difficult cases
in which the basic unregularized method does not yield meaningful results. In case regularization is used,
then both the value of the regularization parameter and the method used to define such value should definitely be quoted,
regardless of the unfolding method or software used.

While most of this information is (or should be) already available to the analyzers
and could therefore be dumped into the main publication document quite easily,
its production might result in a significant amount of additional work, particularly for publications with tens of unfolded distributions;
in cases where the theory community is adamant about the need of using such information,
Collaborations already tend to provide this material---the CMS Collaboration~\cite{Khachatryan:2016mlc} has for example
included the response and covariance matrices in the \textsc{HepData} entry of the publication, to be used by theoreticians for PDF computations.
When the use case is not already explicit or clear, though, the Collaborations tend to not see the value
gained by the additional effort; the theory community can therefore probably help
by proposing concrete ways of making use of any additional information the experiments could provide.

\section{Discussion}

Tools for easy dissemination of useful information are already in place.
Early studies for the creation of public databases have made use of available generator tools such as \textsc{MadAnalysis}~\cite{Dumont:2014tja},
and have evolved into tools like \textsc{Rivet} and \textsc{HepData}, which provide handy ways of sharing most if not all the information discussed above 
even in cases where such information would be too large or require a format unsuitable for the publication text,
There are fora dedicated to the interpretation and recasting of LHC results for BSM studies
such as the LHC Reinterpretation forum~\cite{lhcrecast}.
The latter recently released a preliminary set of recommendations for reinterpreting LHC results for new physics~\cite{Abdallah:2020pec}.

Recommendations for the presentation of LHC results have been developed since the Run I of the LHC~\cite{Kraml:2012sg}
and have been recently updated with proposals toward the definition of an Analysis Description Language for LHC analyses~\cite{Brooijmans:2016vro},
which could improve the situation even more, defining an universal way of sharing information
that would be compatible out of the box for the purpose of later studies.

Sometimes additional material about an analysis is made public by the Collaboration.
Whenever the journal accepts the submission of additional material and publishes it as companion
material to the analysis (perhaps in the journal's page), the Collaboration should follow this practice.
This however sometimes is not possible (the journal may not give the possibility of including material
not contained in the main publication text); Collaborations should then have the possibility of
referencing their own webpages in the publication text as a source of additional material.
Dedicated webpages with additional material for each analysis are already the standard in most experimental Collaboration,
as a way of centralizing the archive of all the produced material; I am simply arguing for the possibility
of citing such pages as a regular reference---possibly with a \textsc{doi} number---in the publication text,
to ensure the highest diffusion also among people that might not be aware of this additional source of material on a given analysis.
The need for such references is highlighted by a comment received during the peer review of the present document,
where one referee seemed unaware that most experimental Collaborations already archive additional material in a
completely independent way from the journals which publish the results.

The level of shared information---even in presence of these tools---is still often insufficient;
I tried in this manuscript to outline the most striking issues that I observed in my daily activities,
without pretending to provide a complete list. I hope this can spark a deeper discussion about such issues
and a higher level of sharing information for the HEP results to come.

\section{Summary}

I have outlined a number of common issues in the reporting of high energy physics results,
mostly focussing on statistical issues.

Taking inspiration from examples from multiboson measurements at ATLAS and CMS,
I have abstracted a certainly non-exhaustive list of the minimal details that should be quoted for each procedure,
outlining that some tools are already readily available and some others are in course of development.
The choice about how much information to share resides ultimately within the Collaborations themselves;
I identified a few areas in which improvement might be substantial,
with the hope that this abstraction might be a guide for sparking a discussion
about reporting results of HEP analyses at the LHC and beyond.